# HD 143811 AB b: A Directly Imaged Planet Orbiting a Spectroscopic Binary in Sco-Cen

Nathalie K. Jones[1,2], Jason J. Wang[1,2], Eric L. Nielsen[3], Robert J. De Rosa[4], Anne E. Peck[3], William Roberson[3], Jean-Baptiste Ruffio[5], Jerry W. Xuan[6,7,41], Bruce A. Macintosh[6], S. Mark Ammons[8], Vanessa P. Bailey[9], Travis S. Barman[10], Joanna Bulger[11], Eugene Chiang[12], Jeffrey K. Chilcote[13], Gaspard Duchêne[12,14], Thomas M. Esposito[12,15], Michael P. Fitzgerald[16], Katherine B. Follette[17], Stephen Goodsell[18,19], James R. Graham[12], Alexandra Z. Greenbaum[20], Pascale Hibon[4], Patrick Ingraham[21], Paul Kalas[12,15,22], Quinn M. Konopacky[5], Michael C. Liu[11], Franck Marchis[15], Jérôme Maire[5], Christian Marois[23,24], Brenda Matthews[23,24], Dimitri Mawet[9,25], Stanimir Metchev[26], Maxwell A. Millar-Blanchaer[27], Rebecca Oppenheimer[28], David W. Palmer[8], Jenny Patience[29], Marshall D. Perrin[30], Lisa Poyneer[8], Laurent Pueyo[30], Abhijith Rajan[30], Julien Rameau[14,31], Fredrik T. Rantakyrö[19], Bin Ren[32], Aniket Sanghi[33,40], Dmitry Savransky[34], Adam C. Schneider[35,36], Anand Sivaramakrishnan[30], Adam J. R. W. Smith[3], Inseok Song[37], Remi Soummer[30], Sandrine Thomas[21], Kimberly Ward-Duong[38], and Schuyler G. Wolff[39]

[1] Department of Physics and Astronomy, Northwestern University, 2145 Sheridan Rd., Evanston, IL 60208-3112, USA; nathaliejones2028@u.northwestern.edu
[2] Center for Interdisciplinary Exploration and Research in Astrophysics, 1800 Sherman Ave., Northwestern University, Evanston, IL 60201, USA
[3] Department of Astronomy, New Mexico State University, P.O. Box 30001, MSC 4500, Las Cruces, NM 88003, USA
[4] European Southern Observatory, Alonso de Córdova 3107, Vitacura, Casilla 19001, Santiago, Chile
[5] Department of Astronomy & Astrophysics, University of California San Diego, La Jolla, CA, USA
[6] Department of Astronomy and Astrophysics, UC Santa Cruz, Santa Cruz, CA 95064, USA
[7] Department of Earth, Planetary, and Space Sciences, University of California, Los Angeles, CA 90095, USA
[8] Lawrence Livermore National Laboratory, 7000 East Ave., Livermore, CA 94550, USA
[9] Jet Propulsion Laboratory, California Institute of Technology, 4800 Oak Grove Dr., Pasadena, CA 91109, USA
[10] Lunar and Planetary Lab, University of Arizona, Tucson, AZ 85721, USA
[11] Institute for Astronomy, University of Hawai'i, 2680 Woodlawn Dr., Honolulu, HI 96822, USA
[12] Department of Astronomy, 501 Campbell Hall, University of California Berkeley, Berkeley, CA 94720-3411, USA
[13] Department of Physics and Astronomy, University of Notre Dame, 225 Nieuwland Science Hall, Notre Dame, IN 46556, USA
[14] University of Grenoble Alpes, CNRS, IPAG, F-38000 Grenoble, France
[15] SETI Institute, Carl Sagan Center, 339 Bernardo Ave., Ste. 200, Mountain View, CA 94043, USA
[16] Department of Physics & Astronomy, University of California, Los Angeles, CA 90095, USA
[17] Physics and Astronomy Department, Amherst College, 25 East Dr., Amherst, MA 01002, USA
[18] Department of Physics, Durham University, Stockton Rd., Durham DH1, UK
[19] Gemini Observatory, Casilla 603, La Serena, Chile
[20] IPAC, Mail Code 100-22, Caltech, 1200 E. California Blvd., Pasadena, CA 91125, USA
[21] Vera C. Rubin Observatory, 950 N Cherry Ave., Tucson AZ 85719, USA
[22] Institute of Astrophysics, FORTH, GR-71110 Heraklion, Greece
[23] Herzberg Astronomy and Astrophysics, National Research Council of Canada, 5071 West Saanich Rd., Victoria, BC V9E 2E7, Canada
[24] Department of Physics & Astronomy, University of Victoria, 3800 Finnerty Rd., Victoria, BC V8P 5C2, Canada
[25] Department of Astronomy, California Institute of Technology, Pasadena, CA 91125, USA
[26] Department of Physics & Astronomy, Institute for Earth and Space Exploration, The University of Western Ontario, London, ON N6A 3K7, Canada
[27] Department of Physics, University of California, Santa Barbara, CA 93106, USA
[28] American Museum of Natural History, Department of Astrophysics, Central Park West at 79th St., New York, NY 10024, USA
[29] School of Earth and Space Exploration, Arizona State University, Tempe, AZ 85287, USA
[30] Space Telescope Science Institute, 3700 San Martin Dr., Baltimore, MD 21218, USA
[31] Trottier Institute for Research on Exoplanets, Université de Montréal, Département de Physique, C.P. 6128 Succ. Centre-ville, Montréal, QC H3C 3J7, Canada
[32] Observatoire de la Côte d'Azur, 96 Bd de l'Observatoire, 06304 Nice, France
[33] Cahill Center for Astronomy and Astrophysics, California Institute of Technology, 1200 E. California Blvd., MC 249-17, Pasadena, CA 91125, USA
[34] Sibley School of Mechanical and Aerospace Engineering, Cornell University, Ithaca, NY 14853, USA
[35] United States Naval Observatory, Flagstaff Station, 10391 West Naval Observatory Rd., Flagstaff, AZ 86005, USA
[36] Department of Physics and Astronomy, George Mason University, MS3F3, 4400 University Dr., Fairfax, VA 22030, USA
[37] Department of Physics and Astronomy, University of Georgia, Athens, GA 30602, USA
[38] Department of Astronomy, Smith College, Northampton, MA 01063, USA
[39] Steward Observatory, University of Arizona, Tucson, AZ 85721, USA

## Abstract

We present confirmation of HD 143811 AB b, a substellar companion to spectroscopic binary HD 143811 AB through direct imaging with the Gemini Planet Imager (GPI) and Keck NIRC2. HD 143811 AB was observed as a part of the GPI Exoplanet Survey in 2016 and 2019 and is a member of the Sco-Cen star formation region. The exoplanet is detected

[40] NSF Graduate Research Fellow.
[41] 51 Pegasi b Fellow.

~430 mas from the host star by GPI. With two GPI epochs and one from Keck/NIRC2 in 2022, we confirm through common proper motion analysis that the object is bound to its host star. We derive an orbit with a semimajor axis of $64^{+32}_{-14}$ au and eccentricity $0.23^{+0.24}_{-0.16}$. Spectral analysis of the GPI *H*-band spectrum and NIRC2 $L'$ photometry provides additional proof that this object is a substellar companion. We compare the spectrum of HD 143811 AB b to PHOENIX stellar models and Exo-Radioactive-Convective Equilibrium Model (REM) exoplanet atmosphere models and find that Exo-REM models provide the best fits to the data. From the Exo-REM models, we derive an effective temperature of $1042^{+178}_{-132}$ K for the planet and translate the derived luminosity of the planet to a mass of $5.6 \pm 1.1\,M_{\rm Jup}$ assuming hot-start evolutionary models. HD 143811 AB b is the first directly imaged planet around a binary that is not on an ultrawide orbit. Future characterization of this object will shed light on the formation of planets around binary star systems.

## 1. Introduction

The direct imaging of exoplanets is a technically demanding endeavor that requires spatially separating the faint light of the planet from its host star. Since the first detections of exoplanets with direct imaging (G. Chauvin et al. 2004; C. Marois et al. 2008; A. M. Lagrange et al. 2009), this technique allows discovery of gas giant planets down to orbital separations of 3 au (e.g., M. Nowak et al. 2020). Some of the successfully imaged planets are a result of targeted searches, where the gravitational influence of the planet was identified from other observations (e.g., $\beta$ Pic c, M. Nowak et al. 2020; AF Lep b, R. J. De Rosa et al. 2023; K. Franson et al. 2023; D. Mesa et al. 2023; HD 206893 c, S. Hinkley et al. 2023; 14 Her c, D. C. Bardalez Gagliuffi et al. 2025; and Eps Indi b, E. C. Matthews et al. 2024). Many others are discovered through blind surveys of young, nearby stars, where direct imaging instruments are most sensitive to planets (e.g., B. Macintosh et al. 2015; G. Chauvin et al. 2017). By imaging these planets over time, we are able to measure their orbital motion, identify gravitationally stable orbits (e.g., J. J. Wang et al. 2018a), and study how they influence smaller bodies in the system (e.g., M. A. Millar-Blanchaer et al. 2015). Additionally, by directly observing the light of the planets, we can spectroscopically study the emission of their atmosphere in order to measure the temperature (A. Z. Greenbaum et al. 2018), elemental composition (E. Nasedkin et al. 2024), radial velocity (I. A. G. Snellen et al. 2014), and rotation speeds (I. A. G. Snellen et al. 2014).

Nearly all of the directly imaged planets discovered to date are found orbiting single stars.[42] One contributing factor for the lack of planets imaged around binaries is that direct imaging surveys generally disfavor binary stars in their target selection strategy (C. Thalmann et al. 2014). Moreover, binarity is known to strongly effect planetary systems (A. L. Kraus et al. 2016). While difficult to image, planets around binaries can provide an invaluable perspective into the role of dynamics on planetary architecture. There are only a few directly imaged exoplanets known to orbit around an inner binary. HD 106906 b (V. Bailey et al. 2014; L. Rodet et al. 2017), b Cen b (M. Janson et al. 2021) and WISPIT 1bc (R. F. van Capelleveen et al. 2025) are directly imaged planets that orbit at extremely wide separations that make in situ formation through core accretion unlikely. HD 106906 b is thought to have experienced a series of dynamical interactions that resulted in its final orbit to be >700 au from the stellar binary, which orbit each other within 1 au (L. Rodet et al. 2017; M. M. Nguyen et al. 2021). It may also have a high planetary obliquity relative to the system's debris disk, which could be explained by formation through turbulent gravitational instability (M. L. Bryan et al. 2021). For b Cen b, the large projected separation of 550 au combined with a modest eccentricity disfavors strong dynamical interactions and may point to it forming in situ through gravitational instability (M. Janson et al. 2021). There are also a few planetary-mass companions with masses that straddle the boundary between planets and brown dwarfs that orbit around binaries: Ross 458C (B. Goldman et al. 2010; Z. Zhang et al. 2021), SR 12 C (M. Kuzuhara et al. 2011; Y.-L. Wu et al. 2022), 2MASS J01033563-5515561 (AB)b (P. Delorme et al. 2013), and ROXs 42B b (T. Currie et al. 2014; A. L. Kraus et al. 2014). These four objects orbit low-mass stars and, with the exception of 2MASS J01033563-5515561 (AB)b, definitely orbit beyond 100 au. For ROXs 42B b, these reasons, combined with a solar atmospheric composition, point to formation via gravitational instability (J. W. Xuan et al. 2024).

The Gemini Planet Imager (GPI) was a high-contrast imaging instrument on Gemini South (B. Macintosh et al. 2014). GPI was equipped with a high-order adaptive optics system to correct atmospheric turbulence (L. A. Poyneer et al. 2016), an apodized Lyot coronagraph to suppress the glare of the star (R. Soummer et al. 2011), and a near-infrared integral field spectrograph (IFS) for imaging exoplanet systems (J. K. Chilcote et al. 2012). The GPI Exoplanet Survey (GPIES) was a 5 yr survey of young, nearby stars to search for gas giant planets. GPIES discovered the first imaged T dwarf planet, 51 Eri b (B. Macintosh et al. 2015), a brown dwarf orbiting within a debris disk (Q. M. Konopacky et al. 2016), and imaged many other debris disks (T. M. Esposito et al. 2020). The first half of GPIES shows that giant planets are relatively rare beyond 10 au and that their occurrence rate is positively correlated with host star mass (E. L. Nielsen et al. 2019). A small fraction of the stars observed by GPI were known spectroscopic binaries with inferred angular separations between the host star to be <20 mas, since these did not affect the high-contrast performance of GPI (E. L. Nielsen et al. 2019). Furthermore, at the time of the survey, HD 143811 AB had not yet been identified as a binary and was therefore included as a target.

In this work, we present the confirmation of a ~5–6 Jupiter mass planet orbiting the spectroscopic binary at HD 143811 AB at ~60 au. Notably, this object is the first directly imaged exoplanet orbiting an binary at solar system scales. This work is part of the final planet vetting and data analysis by the GPIES team. We note that the planet was also confirmed in an independent analysis by V. Squicciarini et al. (2025) using GPI

[42] Note that some directly imaged systems such as 51 Eri b (B. Macintosh et al. 2015) and HD 135344 Ab (T. Stolker et al. 2025) orbit a single star but have stellar companions at large separations.

**Table 1**
Properties of the HD 143811 AB System

| Parameter | HD 143811 A | HD 143811 B | HD 143811 AB b | References |
|---|---|---|---|---|
| Age (Myr) | $13 \pm 4$ | | ⋯ | (1) |
| Distance (pc) | $136.9 \pm 0.4$ | | ⋯ | (2) |
| Mass ($M_\odot$, $M_{\mathrm{JUP}}$) | $1.30^{+0.03}_{-0.05}$ | $1.15^{+0.03}_{-0.04}$ | $5.6 \pm 1.1$ | (1, 3) |
| $T_{\mathrm{eff}}$ (K) | $6751^{+207}_{-148}$ | $6349^{+122}_{-133}$ | $1042^{+178}_{-132}$ | (1, 3) |
| $a$ (au) | $0.1854^{+0.0014}_{-0.0024}$ | | $64^{+32}_{-14}$ | (1, 3) |
| $e$ | $0.4941 \pm 0.0013$ | | $0.23^{+0.24}_{-0.16}$ | (1, 3) |
| $i$ (deg) | $22.9^{+0.3}_{-0.2}$ or $157.1^{+0.2}_{-0.3}$ | | $38 \pm 16$ | (1, 3) |
| $\omega$ (deg) | $155.96 \pm 0.19$ | $335.96 \pm 0.19$ | $180^{+115}_{-130}$ | (1, 3) |
| $\Omega$ (deg) | ⋯ | | $180^{+130}_{-115}$ | (3) |
| $T_0$ (BJD, yr) | $2458230.946 \pm 0.006$ | | $2110^{+180}_{-52}$ | (1, 3) |
| $P$ (days, yr) | $18.59090 \pm 0.00007$ | | $330^{+280}_{-100}$ | (1, 3) |

**Notes.**
**References**. (1) A. E. Peck et al. 2025, (2) Gaia Collaboration et al. (2023), (3) this work.
[a] In the parentheses, the first unit is for HD 143811 A and HD 143811 B, and the second unit is for HD 143811 AB b.

data and with an additional new epoch from SPHERE. In this Letter, Section 2 summarizes the host binary star properties, Section 3 describes the GPI and NIRC2 data used in this analysis, and Section 4 discusses the confirmation of the planet HD 143811 AB b through both astrometric and spectroscopic analyses.

## 2. HD 143811 AB

HD 143811 AB is a spectroscopic binary (O. V. Zakhozhay et al. 2022; A. Grandjean et al. 2023) in the Sco-Cen star formation region (P. T. de Zeeuw et al. 1999). We present in-depth characterization of this binary in a companion Letter (A. E. Peck et al. 2025) but summarize key results here.

HD 143811 AB is a member of the Sco-Cen star formation region (P. T. de Zeeuw et al. 1999). The Sco-Cen subgroup to which it belongs is somewhat uncertain; P. T. de Zeeuw et al. (1999), M. J. Pecaut et al. (2012), and P. A. B. Galli et al. (2018) find it to be a member of Upper Scorpius (US), though K. L. Luhman & T. L. Esplin (2020) do not. The next most likely subgroup is Upper Centaurus–Lupus (UCL; J. Gagné et al. 2018). Peck et al. use a new system radial velocity and BANYAN Σ (J. Gagné et al. 2018) to produce the following membership probabilities: 69.5% for US, 30.3% for UCL, and 0.2% chance in the field. We assign an age of 13 ± 4 Myr for HD 143811 AB, from ages of Upper Sco of 10 ± 3 Myr and UCL of ∼16 ± 2 Myr, which follow the 2D map of Sco-Cen ages of M. J. Pecaut & E. E. Mamajek (2016).

Based on unresolved photometry and multiepoch high-resolution spectra obtained with FEROS (A. Kaufer et al. 1999), the stellar binary HD 143811 AB b has a period of 18.59098 ± 0.00007 days and a mass ratio of 0.885 ± 0.003. From this analysis, we adopt masses of the two components of $M_A = 1.30^{+0.03}_{-0.05}\ M_\odot$ and $M_B = 1.16^{+0.03}_{-0.04}\ M_\odot$. With only a double-lined spectroscopic binary radial velocity orbit, we cannot currently constrain the full 3D orbit of the binary; however, future orbit monitoring may shed light on the full architecture of this system. Key parameters of the HD 143811 AB system are summarized in Table 1.

To generate a synthetic spectral energy distribution of the unresolved binary, we used the procedure outlined in E. L. Nielsen et al. (2019). A joint evolutionary and atmospheric model was created using a modified version of the MESA Isochrones and Stellar Tracks model grid (J. Choi et al. 2016; A. Dotter 2016) and the ATLAS9 atmospheric model spectra (F. Castelli & R. L. Kurucz 2003). An Markov Chain Monte Carlo (MCMC)–based approach was used to sample the posterior distributions of the masses of each component ($M_1$, $M_2$); the age ($t$), metallicity ([$M$/H]), and parallax ($\varpi$) of the system; and the visual extinction toward the system ($A_V$). Gaussian priors on the age (13 ± 4 Myr; synthesizing the age for the US and UCL subgroups from M. J. Pecaut & E. E. Mamajek 2016), parallax (7.3065 ± 0.0204 mas; Gaia Collaboration et al. 2023), metallicity (−0.05 ± 0.11 dex; E. L. Nielsen et al. 2013), and mass ratio (0.889 ± 0.004) were used. At each step, the predicted flux of the blended system was compared to the Gaia (Gaia Collaboration et al. 2023) and Two Micron All Sky Survey (M. F. Skrutskie et al. 2006) photometry. A synthetic spectrum at GPI's resolution, as well as a synthetic $L$-band flux, were calculated by drawing randomly from the converged chain.

## 3. Observations and Data Analysis

In this Letter, we utilize two data sets from the GPI at Gemini South and one data set from NIRC2 at the W. M. Keck Observatory. We provide more details on the observations from both instruments below.

### 3.1. Gemini/GPI Observations

HR 143811 AB was observed by GPI two times as part of GPIES. Both observations were obtained with the GPI IFS in $H$ band (1.50–1.80 $\mu$m) with a spectral resolution $\Delta\lambda/\lambda \sim 40$ and a field of view of 190 × 190 pixels (2″.7 × 2″.7). The first epoch was taken on 2016 April 30 (UT) (GS-2015B-Q-500) and consists of 36 good exposures of 60 s. The second epoch was taken on 2019 August 11 (UT) (GS-2019A-Q-500) and consists of 45 good exposures of 60 s. The 2016 observation spanned 174°.5 of parallactic angle rotation, while the 2019 observation spanned 174°.8. The DIMM seeing in 2016 was 2″.24 at the beginning of the observation. While a precise measurement on the 2019 seeing conditions is unavailable, the observing team noted poor conditions on that night, leading to a low signal-to-noise ratio (SNR) detection.

Initial data reduction for each epoch was done using the automated data reduction system for GPIES (J. J. Wang et al. 2018b). The details are listed in Appendix A of J. J. Wang et al. (2018b), but we will briefly summarize the steps here. Individual raw 2D data from the IFS were transformed into 3D spectral data cubes ($x$, $y$, $\lambda$) using the GPI Data Reduction Pipeline version 1.6.0 (M. D. Perrin et al. 2014, 2016). The GPI Data Reduction Pipeline also corrects for bad pixels, corrects for distortion (Q. M. Konopacky et al. 2014), and measures the position and flux of the four satellite spots (C. Marois et al. 2006b; A. Sivaramakrishnan & B. R. Oppenheimer 2006) for astrometric and photometric calibration (J. J. Wang et al. 2014). The stellar point-spread function (PSF) is subtracted from each data set using `pyKLIP` (J. J. Wang et al. 2015), a Python implementation of the Karhunen–Loève image projection (KLIP) algorithm, using both angular differential imaging (ADI; M. C. Liu 2004; C. Marois et al. 2006a) and spectral differential imaging (C. Marois et al. 2000). The forward model matched filter (FMMF) planet detection code (J.-B. Ruffio et al. 2017) was then run on each epoch using both an L-type and T-type spectral template. The FMMF flux maps are shown in Figure 1 for each epoch. Using the L-type spectral template, HD 143811 AB b was detected at an SNR of 7.7 in the 2016 epoch and an SNR of 3.5 in the 2019 epoch.

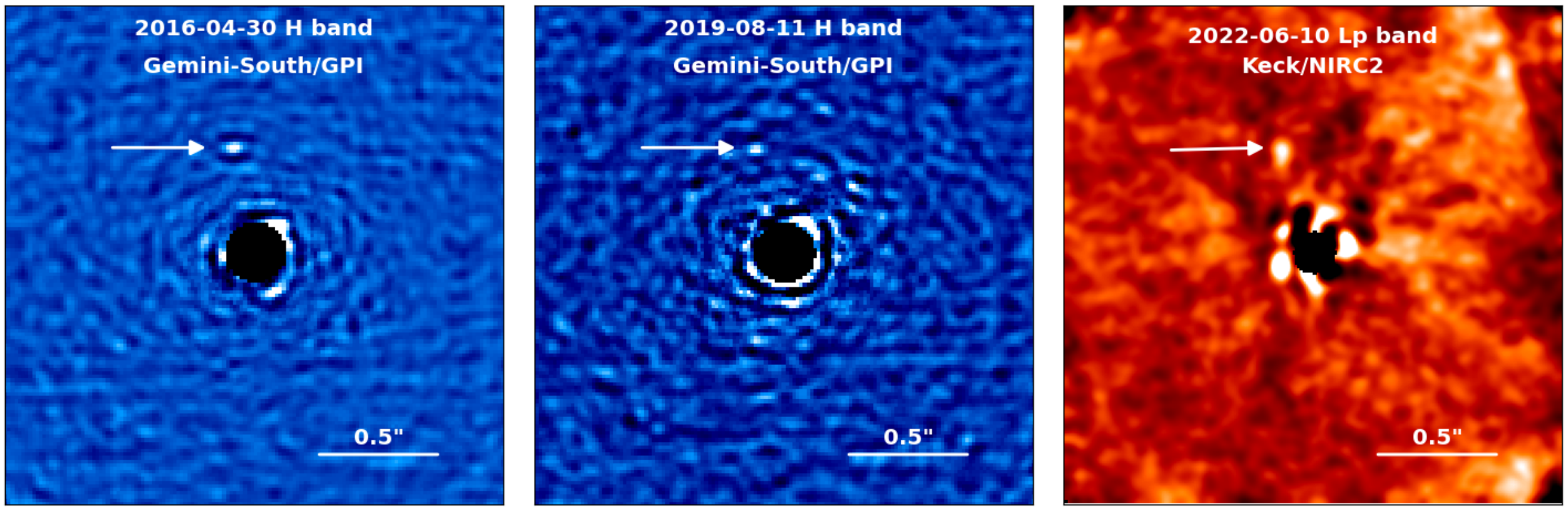

**Figure 1.** Reduced flux maps of HD 143811 AB obtained by Gemini/GPI in 2016 and 2019 (left and center) and Keck/NIRC2 in 2022 (right). These images are 2” × 2”. North is up, and east is left. The star is at the center of the image behind the coronagraph. Arrows indicates the location of the planet. Intensity scales are linear and vary for each image in order to highlight the object.

**Table 2**
Observations and Measurements of HD 143811 AB b

| Date (UT) | 2016 Apr 30 | 2019 Aug 11 | 2022 Jun 10 |
|---|---|---|---|
| Instrument | GPI | GPI | NIRC2 |
| Filter | $H$ | $H$ | $L'$ |
| Total Integration Time (minute) | 36 | 45 | 87.5 |
| $\rho$ (mas) | 429 ± 3 | 429 ± 10 | 421 ± 9 |
| $\theta$ (deg) | 11.9 ± 1.0 | 15 ± 3 | 18.7 ± 1.2 |
| Flux Ratio | $(1.4 \pm 0.3) \times 10^{-5}$ | $0.8^{+0.9}_{-0.5} \times 10^{-5}$ | $(1.8 \pm 0.8) \times 10^{-4}$ |
| Delta Magnitude | 12.1 ± 0.1 | ⋯ | 8.89 ± 0.09 |
| Apparent Magnitude | 19.8 ± 0.2 | ⋯ | 16.6 ± 0.3 |
| Absolute Magnitude | 14.1 ± 0.2 | ⋯ | 10.9 ± 0.3 |
| Flux ($\mathrm{W m^{-2} \mu m^{-1}}$) | $(1.4 \pm 0.2) \times 10^{-17}$ | $0.8^{+0.8}_{-0.5} \times 10^{-17}$ | $(1.3 \pm 0.4) \times 10^{-17}$ |

We also utilize `pyKLIP` to measure the astrometry, obtain the relative flux ratio of the point source, and extract a spectrum. To obtain precise astrometric measurements, we first generate a forward model (FM) template of the planet PSF through the PSF subtraction process (A. M. Lagrange et al. 2010; C. Marois et al. 2010) using the KLIP-FM formalism (L. Pueyo 2016; J. J. Wang et al. 2016) and using the known flux ratio between the satellite spots and the star (R. J. De Rosa et al. 2020a). Then, the posterior of the planet’s location is obtained using Bayesian parameter estimation using the `emcee` package (D. Foreman-Mackey et al. 2013) and Gaussian processes to model the spatially correlated noise in the images (J. J. Wang et al. 2016). To convert from pixel locations to relative astrometry, we use a plate scale of $14.166 \pm 0.007$ mas pixel$^{-1}$ and a north-angle offset of $-0.10 \pm 0.13$ deg (R. J. De Rosa et al. 2020b). We also include a 0.05 pixel uncertainty in locating the central binary (J. J. Wang et al. 2014). The astrometry for both epochs is tabulated in Table 2. Following A. Z. Greenbaum et al. (2018), KLIP-FM is also used to extract an $H$-band spectrum from the 2016 April 30 epoch, because it is the best detection. The error on the extracted spectrum is measured by injecting simulated point sources into the data at the same distance from the star as the real planet, reextracting the simulated source spectra, and using the standard deviation of the reextracted spectra across all injected sources as the error in the extracted spectrum. Then, we obtain the final planet spectrum by normalizing against the stellar spectral energy distribution (SED) model described in Section 2. Like other GPI spectra (e.g., A. Rajan et al. 2017), the spectra exhibits a high degree of correlated noise, with the scatter between neighboring data points being much smaller than the calculated uncertainties.

### *3.2. Keck/NIRC2 Observations*

HD 143811 AB was observed on 2022 June 10 (UT) with Keck/NIRC2 in $L'$ band (3.426–4.126 $\mu$m) using the infrared pyramid wave front sensor to correct for atmospheric turbulence (C. Z. Bond et al. 2018) but not using a coronagraph. One hundred seventy-five exposures were taken, with each exposure comprising of 60 coadds of 0.5 s integration times and utilizing only a 264 × 256 subwindow of the detector (∼2.″6 × 2.″5 field of view). A total of 87.5 minutes of total integration time on target were obtained. Additional shorter exposures comprising of 60 coadds of 0.1 s exposures were taken of the star for flux calibration. The data were taken in vertical angle mode, enabling ADI to be used. The observations spanned a total of 39.°5 of parallactic angle rotation. The average DIMM seeing was 0.″43 during the observations. We note that there is another epoch of Keck/NIRC2 data in $L'$ band taken on 2022 April 29 with the vortex coronagraph. However, the star appears significantly offset from the center of the vortex coronagraph in that epoch, making measurements such as relative astrometry biased. Thus, we did not include that epoch in our analysis.

We performed initial preprocessing of the data using a general NIRC2 pipeline originally developed for the vortex coronagraph (W. J. Xuan et al. 2018; G. Ruane et al. 2019). First, we corrected bad pixels and flat-field effects in each image. Then, we subtracted the thermal background from the sky and instrument using principal component analysis, similar to what was done in R. Galicher et al. (2011). Afterwards, the star position in each frame was measured with a 2D Gaussian fit to each frame. We then performed stellar PSF subtraction using `pyKLIP` to remove the glare of the star from this preprocessed image sequence. For each science frame, we used all other science frames where the planet would have moved by at least 4 pixels due to ADI to build the 10 principal components that were used to model the stellar PSF off the glare of the star. The resulting images are mean combined, and a Gaussian-smoothed image is shown in Figure 1. The planet is detected at an SNR of 3.9.

To measure the relative astrometry and flux ratio of the planet, we followed a similar procedure to the GPI data. We forward modeled the PSF of the point source through KLIP using the 0.1 s exposures of the star and the KLIP-FM framework, and we performed PSF fitting using MCMC sampling with the `emcee` package to infer the detector position and flux ratio of the object following the procedure. To estimate the uncertainty of the planet parameters, we injected simulated point sources of the same brightness and distance from the star into the data at other azimuthal positions, performed the same PSF fitting routine, and used the spread in the measured properties of the simulated planets as the uncertainty in the detector position and flux ratio of the detected planet. To convert from detector position to relative astrometry, we used the NIRC2 plate scale of 9.971 ± 0.004 mas pixel$^{-1}$ and north-angle offset of 0.°262 ± 0.°020 (M. Service et al. 2016). Then, we added an additional 0.°118 ± 0.°006 north-angle offset that was found for the pyramid wave front sensor mode of NIRC2 (S. Walker et al. 2025) and conservatively assumed a star center uncertainty of 0.5 pixels. We did not apply a distortion correction since the differential distortion across a 500 mas radius circle centered around the star is smaller than the 0.1 pixel uncertainty on the distortion solution (M. Service et al. 2016). To calculate the $L'$ flux of the planet, we scaled it to the star's $L'$ photometry using the stellar SED model in Section 2. The astrometry and photometry are tabulated in Table 2.

## 4. Properties of HD 143811 AB b

### *4.1. Companionship and Orbit*

We confirm that the close detected object is a common proper motion companion using the method described in R. J. De Rosa et al. (2015). Gaia DR3 proper motion and parallax are used to predict the motion of a stationary background object, based on the relative astrometry of the companion at the original epoch of detection (see Figure 2). We also use the Orbits for the Impatient (OFTI) algorithm (S. Blunt et al. 2017) to generate plausible orbit tracks from this first epoch, assuming a total system mass of 2.46 $M_\odot$. The separation measurements disagree with the background track at $\gtrsim 20\sigma$ and are much more consistent with a mostly unchanging separation over the time span of the observations. As a result, we conclude at high confidence that HD 143811 AB b is not a stationary background object.

The relative astrometry is consistent with orbital motion, and we run an OFTI fit on the three epochs of data, assuming a parallax of 7.31 ± 0.02 mas and a total mass of 2.46 ± 0.07 $M_\odot$. Other priors are taken to be uniform in $\log(a)$, eccentricity, $\cos(i)$, argument of periastron, position angle of nodes, and epoch of periastron passage. We find orbital parameters consistent with a moderately inclined orbit, with semimajor axis, eccentricity, inclination angle, and period of $a = 64^{+32}_{-14}$ au, $e = 0.23^{+0.24}_{-0.16}$ (68% of orbits have $e < 0.34$), $i = 38° \pm 16°$, and $P = 330^{+280}_{-100}$ yr, respectively. To determine if there is any significant dynamical mass constraints, we perform an additional orbit fit that incorporates the differential proper motions from the Hipparcos–Gaia Catalog of Accelerations (T. D. Brandt 2021) assuming all of the detected astrometric acceleration is from the HD 143811 AB b, but we were not able to exclude any planet masses between 1 and 100 $M_{\rm Jup}$. We note that while there is a moderate signature of an astrometric acceleration in HGCA ($\chi^2 = 11.3$; T. D. Brandt 2021), there could be biases in the reported proper motions due to the stellar orbit (e.g., see discussion in R. J. De Rosa et al. 2019).

### *4.2. Spectral Analysis*

We performed spectral comparisons in order to assess and characterize the nature of the object. As discussed in E. L. Nielsen et al. (2017), objects can be ruled out as background stars or cool substellar companions based on the spectrum. A catalog of reference spectra was assembled using the PHOENIX synthetic stellar model grid (T. O. Husser et al. 2013) and Exo-Radioactive-Convective Equilibrium Model (REM) exoplanet atmosphere model grid, which includes both clouds and disequilibrium chemistry in its self-consistent model (B. Charnay et al. 2018). We access the Exo-REM grid using the `species` package (T. Stolker et al. 2020). The final comparison sample is 842 templates from PHOENIX and 9575 from Exo-REM. PHOENIX grids for range $T_{\rm eff}$ = 2300–12000 K are utilized. Exo-REM parameters cover a range of $T_{\rm eff}$ = 400–2000 K, $\log(g)$ = 3.0–5.0, metallicity Fe/H = −0.5–1.0, and C/O = 0.1–0.8. The grid spectra are then blurred to the resolution of GPI using a Gaussian filter.

These data were used in a custom $\chi^2$ goodness of fit routine to determine the best-fitting model. We performed the fit for the GPI spectrum alone and, additionally, for the GPI spectrum plus the NIRC2 $L'$ photometry, shown in Figure 3. We do not account for correlated error in the $\chi^2$ analysis. The GPI-only fit results in a spread of preferred effective temperatures (∼660–1100 K). While Exo-REM yields slightly lower $\chi^2$, the PHOENIX grids also fit the data well, with a maximum $\chi^2 \sim 3.4$ (top panel of Figure 4). However, adding the Keck/NIRC2 data to the spectrum shows that the Exo-REM grid is strongly preferred (bottom panel of Figure 4). The rejection of PHOENIX stellar templates provides further evidence that our detection is not a background star but rather a substellar companion. The best-fit ($\chi^2 = 3.31$) model parameters of the GPI + Keck fit are effective temperature, $T_{\rm eff} = 1050$ K, surface gravity, $\log(g) = 4.5$, metallicity 10× solar ([Fe/H] = 1.0), and C/O ratio equal to 0.35, although we note that log(g), [Fe/H], and C/O ratio are unconstrained, as we discuss in the next paragraph. This model is plotted against the candidate spectrum in Figure 3.

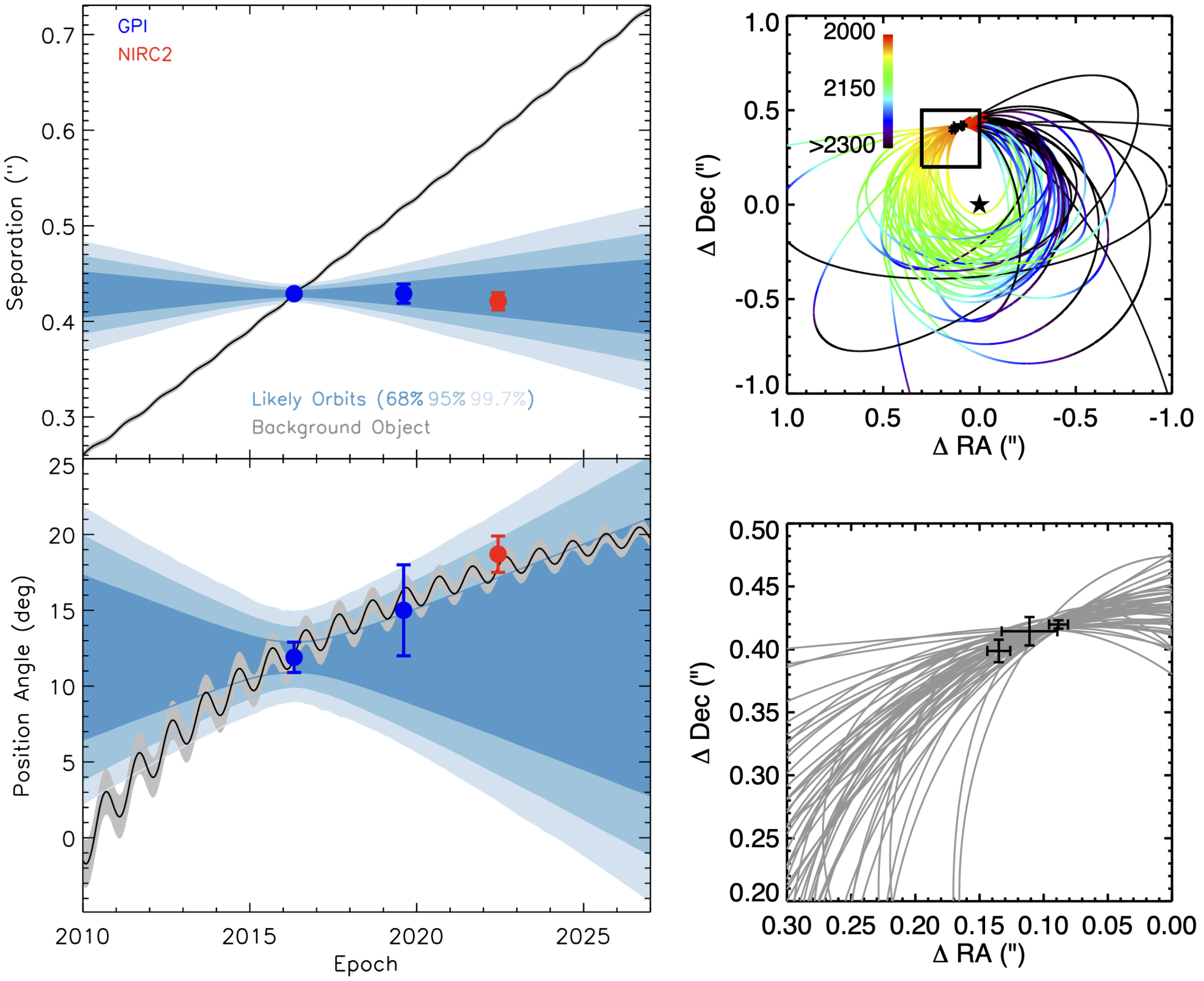

**Figure 2.** (Left) Astrometric measurements for HD 143811 AB b as a function of observational epoch. Points in blue are from GPI, and red denotes Keck NIRC2. The gray lines show the path of a stationary background object, while the blue regions show the paths of possible orbits for a planetary-mass bound companion. With these three epochs, HD 143811 AB b is confirmed to be a bound substellar companion at high confidence. (Top right) Orbit fits to the full relative astrometry data set, with each orbit representing a random draw from the posterior and color-coded by the epoch. (Bottom right) The same orbit tracks but zoomed in to the available data. Orbit fits are consistent with an $e \lesssim 0.5$ orbit with moderate inclination and an orbital period of ~300 yr.

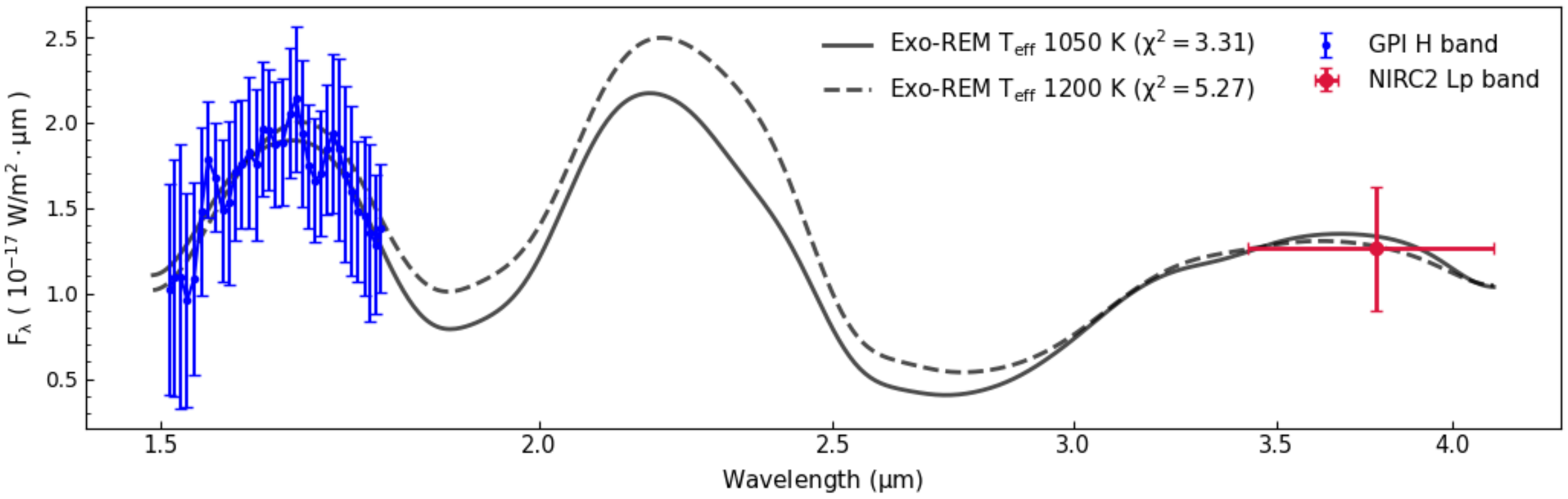

**Figure 3.** Spectrum of HD 143811 AB b from GPI data (blue, $H$ band) and Keck (red, $L'$ band). We also plot the best-fit Exo-REM model (solid gray; $\chi^2 = 3.31$) as well as a second Exo-REM model with a $T_{\rm eff}$ 150 K higher (dashed gray; $\chi^2 = 5.27$).

For a more precise characterization of the planet's atmospheric properties, we fit the GPI $H$-band spectrum and NIRC2 $L'$ photometry to the Exo-REM atmospheric models to estimate the range of likely atmospheric parameters. To account for correlated noise in the GPI spectra, we also fit a Gaussian process to model the correlated noise behavior following the procedure described in J. J. Wang et al. (2020). We sampled the posterior using the nested sampling algorithm (J. Skilling 2004, 2006) with multiple bounding ellipsoids (F. Feroz et al. 2009) from `dynesty` (J. S. Speagle 2020; S. Koposov et al. 2024). The data are sufficient to constrain the effective temperature to $1042^{+178}_{-132}$ K and the radius to $1.7^{+0.7}_{-0.4}$ $R_{\rm Jup}$. The log($g$), metallicity, and C/O ratio are all unconstrained, with allowed values across the entire parameter space of the model grid. From the constrained effective temperature and radius, we derive a planet luminosity of $3.3^{+0.8}_{-0.6} \times 10^{-5} L_\odot$ ($\log_{10}(L/L_\odot) = -4.49^{+0.10}_{-0.09}$). Note that effective temperature and radius are negatively correlated in our posteriors, resulting in a luminosity that is more constrained than the radius itself. Using the age of the system of 13 ± 4 Myr and hot-start evolutionary models from I. Baraffe et al. (2003), we use our computed luminosity to derive a planet mass of 5.6 ± 1.1 $M_{\rm Jup}$. The mass posterior is similar if the system is instead assumed to be part of Upper Sco (10 ± 3 Myr) or UCL (16 ± 2 Myr; M. J. Pecaut & E. E. Mamajek 2016): $4.9^{+0.8}_{-0.9}$ $M_{\rm Jup}$ and $6.2^{+0.8}_{-0.6}$ $M_{\rm Jup}$, respectively. For our adopted age of 13 ± 4 Myr, we also infer a radius of 1.41 ± 0.03$R_{\rm Jup}$ from evolutionary models, consistent with the radius inferred from the Exo-REM fits.

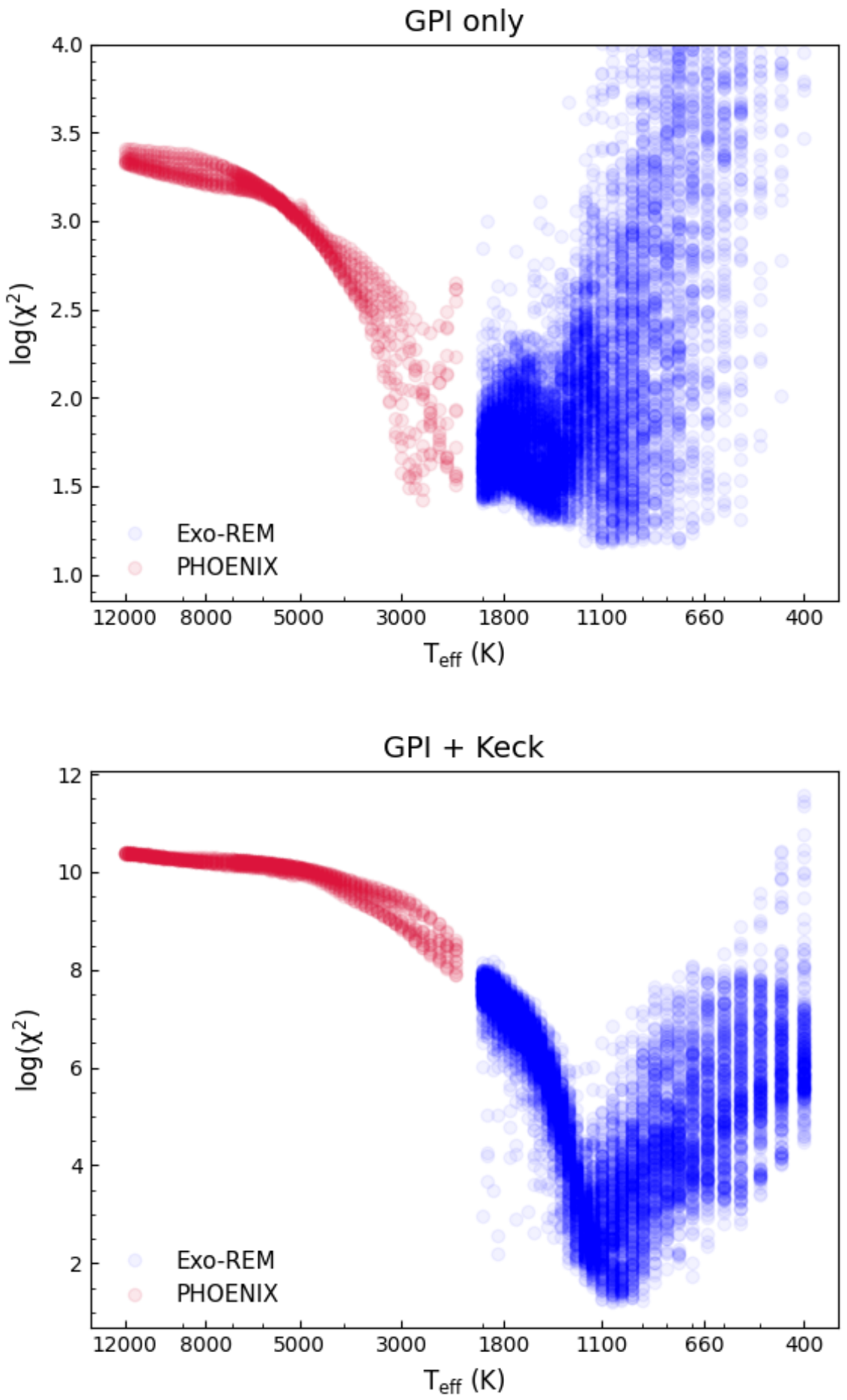

**Figure 4.** Resulting $\chi^2_\nu$ as a function of $T_{\rm eff}$ from spectral comparison of HD 143811 AB b with comparison library. Fit values using the PHOENIX stellar models are shown in red, while the Exo-REM comparisons are shown in blue. Note the different $y$-axes. Both fits prefer the Exo-REM models over PHOENIX, though the addition from Keck focuses the fit to ∼1100 K.

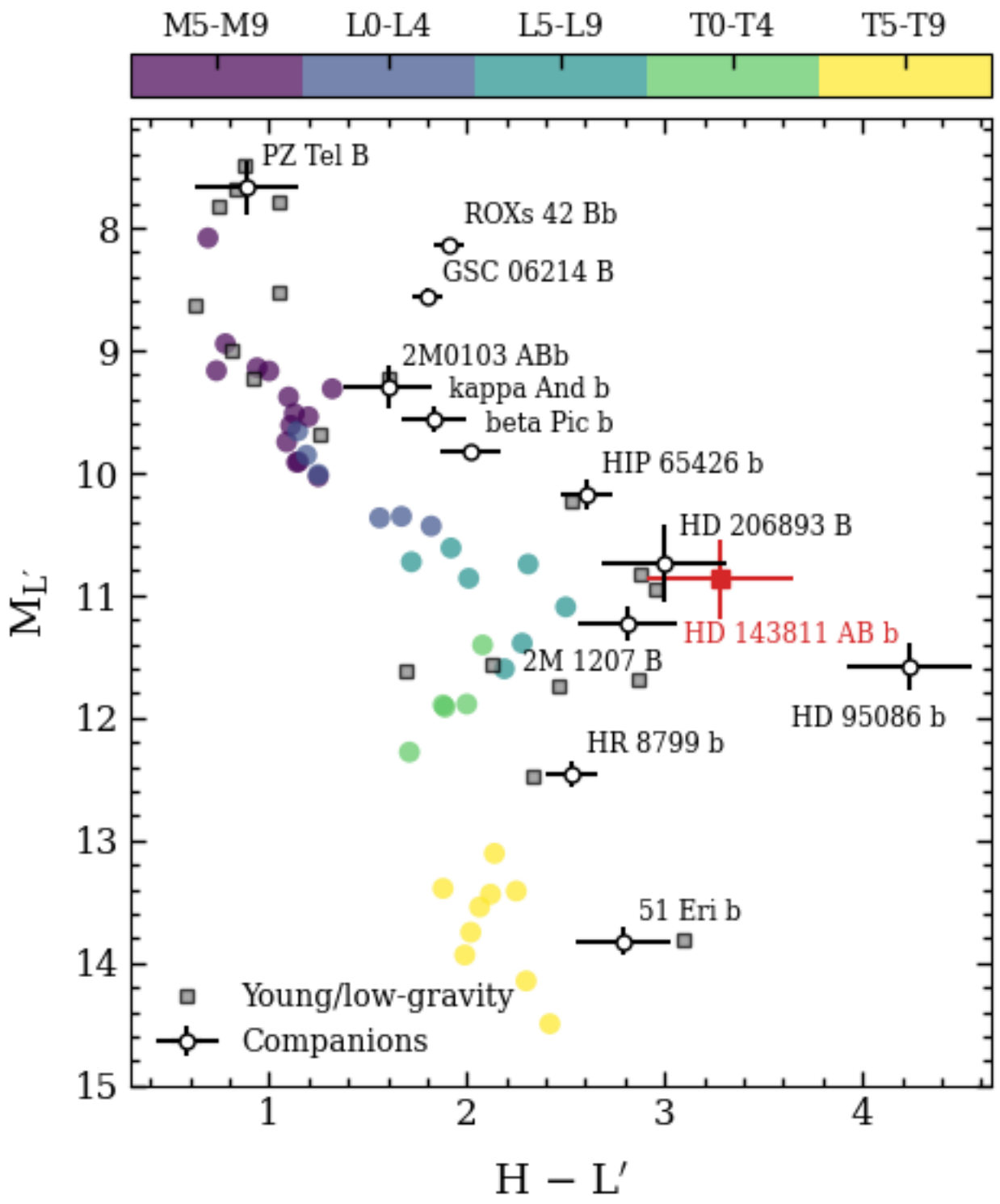

**Figure 5.** Color–magnitude diagram showing HD 143811 AB b (red square) with directly imaged companions, field objects (colored circles), and young/low-gravity objects (squares), which is accessed using `species` (T. Stolker et al. 2020). The color of the field objects varies according to spectral type (color bar is above the plot).

### *4.3. Color–Magnitude Diagram*

Further information on the nature of HD 143811 AB b may be gleaned by adding HD 143811 AB b to a color–magnitude diagram (CMD) with other directly imaged companions and brown dwarfs. In Figure 5, we show a color–magnitude diagram produced with `species`. Colors and magnitudes of field and low-gravity objects are sourced from the UltracoolSheet (W. M. J. Best et al. 2024), whereas companion properties are included within the `species` package. We calculate $H$ and $L'$ magnitudes for HD 143811 AB b and derive $H_{GPI} - L'_{NIRC2} = 3.3 \pm 0.4$ mag, using the 2016 GPI epoch and the NIRC2 data. We do not calculate $H$ magnitude using the 2019 observation since the contrast has large error bars. The absolute magnitude $M_L = 10.87 \pm 0.32$ of HD 143811 AB b is consistent with late L-type or early T-type field objects.

Proximity to other directly imaged companions on a CMD may help infer properties such as spectral type. The exoplanet HD 143811 AB b is located on the CMD near two other directly imaged companions, HD 206893 B (J. Milli et al. 2017) and 2M 1207 B (G. Chauvin et al. 2004). HD 206893 B was also imaged by GPIES and is one of the reddest substellar companions known (S. Hinkley et al. 2023). The GPI observations are presented in K. Ward-Duong et al. (2021) along with discussion of a few sources for the dusty red color, with atmospheric dust in the form of high-altitude aerosols as the most viable. They find a spectral type of L6 ± 2 for HD 206893 B. HD 143811 AB b is also spatially located near 2M 1207 B, which has spectral type L0 ± 0.5 (K. L. Luhman et al. 2023). Lastly, the planet is located near another exceptionally red object, HD 95086 b, which is also considered as a solar system analog (J. Rameau et al. 2013). G. Chauvin et al. (2018) find that spectral types L7 − L9 fit their observations of HD 95086 b. Obtaining additional spectra of HD 143811 AB b may reveal a similarly dusty, red nature.

## 5. Conclusions

HD 143811 AB b is a substellar companion to a spectroscopic binary in the Sco-Cen star formation region. Our conclusions are summarized below:

1. Using the two epochs of GPI data in $H$ band and additional Keck $L'$ data, we rule out the object as a stationary background source using common proper motion.
2. Spectral comparisons of the extracted spectra to stellar and exoplanet atmosphere models additionally confirms that this object is a physically bound exoplanet companion. Fitting the spectrum of HD 143811 AB b to PHOENIX stellar models yields very large values of $\chi^2$, while Exo-REM exoplanet atmosphere models fit our data well (best-fit $\chi^2 = 3.31$).
3. Orbit fitting is performed using OFTI using the three relative astrometry measurements. We find a semimajor axis of $a = 64^{+32}_{-14}$ au and $e = 0.23^{+0.24}_{-0.16}$.
4. We perform spectral fitting of HD 143811 AB b to Exo-REM atmosphere models and derive an effective temperature of $1042^{+178}_{-132}$ K and a planet luminosity of $3.3^{+0.8}_{-0.6} \times 10^{-5} L_{\odot}$. Using hot-start evolutionary models, we derive a planet mass of $5.6 \pm 1.1\, M_{\rm Jup}$.

Most directly imaged exoplanets have been discovered around single stars, not binaries, making HD 143811 AB b an interesting individual system for follow-up characterization. In fact, HD 143811 AB b is the closest-in exoplanet imaged around a binary and is the first imaged exoplanet orbiting a binary at solar system scales. Follow-up observations that better pinpoint the orbit of both HD 143811 AB b and of the binary will help unveil the dynamical architecture of this three-body system, including further constraint on the orbital separation. HD 143811 AB b will be an interesting test case to probe atmospheric signatures of planet formation and assess whether this planet in a binary star system is consistent with the predictions of core accretion. This new planet is also part of a larger sample of planets imaged by the GPIES. Folding this planet and vetting the remaining candidates from GPIES will help improve our constraints on the frequency of giant planets beyond 10 au orbital separations. Future observations with upgraded direct imaging instruments, such as the GPI 2.0 (J. Chilcote et al. 2024; C. Marois et al. 2024), hopefully can discover even more planets around binaries for characterization.

## Acknowledgments

We thank the anonymous referee for helpful suggestions in preparing the final version of this Letter.

N.J. acknowledges the BOBA group for their help and support throughout the preparation and publication of this project. Additionally, N.J. would like to acknowledge and thank their *kūpuna*, *ʻohana*, and the *ʻāina* of Hawaiʻi Island on which part of this work took place.

J.J.W. acknowledges support from the Alfred P. Sloan Foundation.

The Gemini Observatory is operated by the Association of Universities for Research in Astronomy, Inc., under a cooperative agreement with the NSF on behalf of the Gemini partnership: the National Science Foundation (United States), the National Research Council (Canada), CONICYT (Chile), the Australian Research Council (Australia), Ministério da Ciência, Tecnologia e Inovação (Brazil), and Ministerio de Ciencia, Tecnología e Innovación Productiva (Argentina).

This research used resources of the National Energy Research Scientific Computing Center, a DOE Office of Science User Facility supported by the Office of Science of the U.S. Department of Energy under Contract No. DE-AC02-05CH11231. This work used the Extreme Science and Engineering Discovery Environment (XSEDE), which is supported by National Science Foundation grant No. ACI-1548562.

Based on observations collected at the European Organisation for Astronomical Research in the Southern Hemisphere under ESO program(s) 0101.A-9012(A), 0107.A-9004(A), 0109.A-9014 (A). This research has made use of the SIMBAD database, CDS, Strasbourg Astronomical Observatory, France. This research has made use of the VizieR catalog access tool, CDS, Strasbourg Astronomical Observatory, France (F. Ochsenbein et al. 2000).

This work has made use of data from the European Space Agency (ESA) mission Gaia (https://www.cosmos.esa.int/gaia), processed by the Gaia Data Processing and Analysis Consortium (DPAC, https://www.cosmos.esa.int/web/gaia/dpac/consortium). Funding for the DPAC has been provided by national institutions, in particular the institutions participating in the Gaia Multilateral Agreement.

This research was funded in part by the Gordon and Betty Moore Foundation through grant GBMF8550 to M.L.

This material is based on work supported by the National Science Foundation Graduate Research Fellowship under grant No. 2139433 to A.S.

This work was performed under the auspices of the U.S. Department of Energy by Lawrence Livermore National Laboratory under Contract DE-AC52-07NA27344 to S.M A.

This work has benefited from The UltracoolSheet at http://bit.ly/UltracoolSheet, maintained by Will Best, Trent Dupuy, Michael Liu, Aniket Sanghi, Rob Siverd, and Zhoujian Zhang, and developed from compilations by T. J. Dupuy & M. C. Liu (2012), T. J. Dupuy & A. L. Kraus (2013), N. R. Deacon et al. (2014), M. C. Liu et al. (2016), W. M. J. Best et al. (2018), W. M. J. Best et al. (2021), A. Sanghi et al. (2023), and A. C. Schneider et al. (2023).

*Facilities:* Keck:II (NIRC2), Gemini:South (GPI).

*Software*: GPI Data Reduction Pipeline (M. D. Perrin et al. 2014, 2016), `pyKLIP` (J. J. Wang et al. 2015), `VIP` (C. A. Gomez Gonzalez et al. 2017; V. Christiaens et al. 2023), `emcee` (D. Foreman-Mackey et al. 2013), `dynesty` (J. S. Speagle 2020; S. Koposov et al. 2024), `species` (T. Stolker et al. 2020), `astropy` (Astropy Collaboration et al. 2013, 2018, 2022), `numpy` (C. R. Harris et al. 2020), `matplotlib` (J. D. Hunter 2007).

## Appendix A
## Orbit Fit

Included below are plots of orbital motion (Figure A1) and posteriors from OFTI (Figure A2). We also include a table of posteriors and best-fit orbital parameters (Table A1).

**Table A1**
Posteriors and Best-fitting Orbital Parameters of HD 143811 AB b

| Parameter | Median | 68% CI | 95% CI | 99.7% CI | Min. $\chi^2$ | Max Likelihood |
|---|---|---|---|---|---|---|
| Semimajor axis $a$ (au) | 63.881 | 50.345–96.412 | 39.354–202.632 | 33.519–628.152 | 50.764 | 59.547 |
| Eccentricity $e$ | 0.231 | 0.069–0.470 | 0.010–0.714 | 0.001–0.892 | 0.205 | 0.016 |
| Inclination Angle $i$ (deg) | 37.543 | 20.936–53.422 | 7.999–67.826 | 1.941–79.197 | 3.964 | 40.225 |
| Argument of Periastron $\omega$ (deg) | 179.950 | 53.202–295.117 | 8.247–351.328 | 0.519–359.478 | 140.979 | 263.654 |
| Position Angle of Nodes $\Omega$ (deg) | 180.230 | 66.397–311.038 | 6.428–353.251 | 0.393–359.608 | 18.763 | 357.092 |
| Epoch of Periastron Passage $T_0$ (yr) | 2112.574 | 2061.066–2287.424 | 2010.707–2932.706 | 2000.706–7383.107 | 2104.960 | 2210.992 |
| Period $P$ (yr) | 326.347 | 228.138–605.062 | 157.633–1843.184 | 123.828–10034.210 | 238.791 | 284.978 |
| Total Mass $M_{\rm tot}$ ($M_\odot$) | 2.451 | 2.381–2.520 | 2.311–2.591 | 2.241–2.660 | 2.294 | 2.600 |

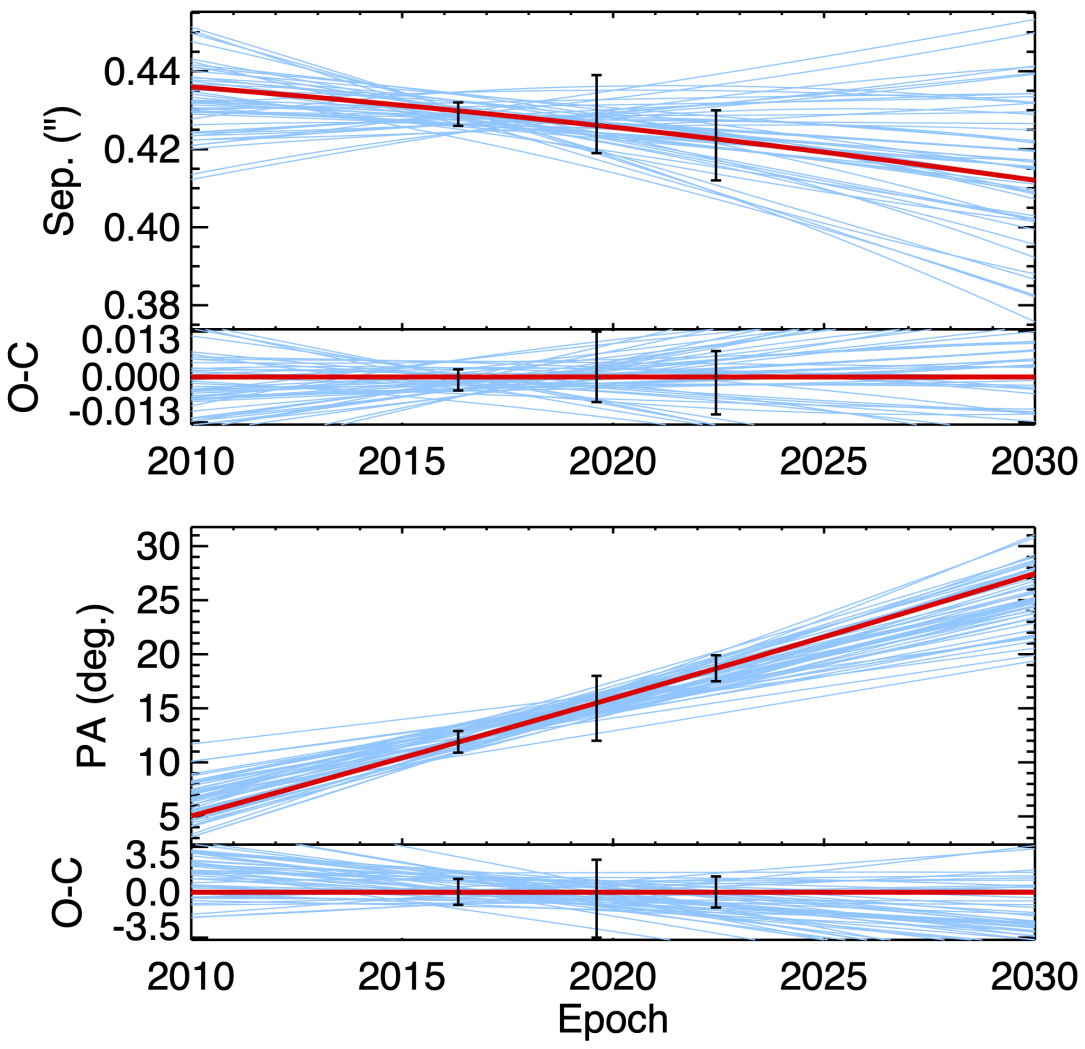

**Figure A1.** Orbital motion from the OFTI fit to the relative astrometry of HD 143811 AB b. Over the next decade, separation is expected to decrease slightly, while the position angle increases (counterclockwise motion).

Posteriors for argument of periastron ($\omega$) and position angle of nodes ($\Omega$) are symmetric about 180° given the standard degeneracy for orbit fit to relative astrometry alone. While there are local maxima in the posterior (e.g., ∼80° and ∼260° for $\omega$), the posterior never reaches 0 probability for any value of either angle. There are, however, combinations of the two that are less favored compared to others, as shown in the covariance plot in Figure A2. As a result, neither the 1D posterior on these two parameters ($\omega = 180^{+115}_{-130}{}^\circ$and $\Omega = 180^{+130}_{-115}{}^\circ$) nor the 180° folded posterior ($\omega = 80^{+70}_{-50}{}^\circ$and $\Omega = 110^{+50}_{-80}{}^\circ$) are particularly informative. We therefore also include in Table A1the minimum-$\chi^2$ and maximum likelihood orbit parameters to aid in choosing a representative orbit.

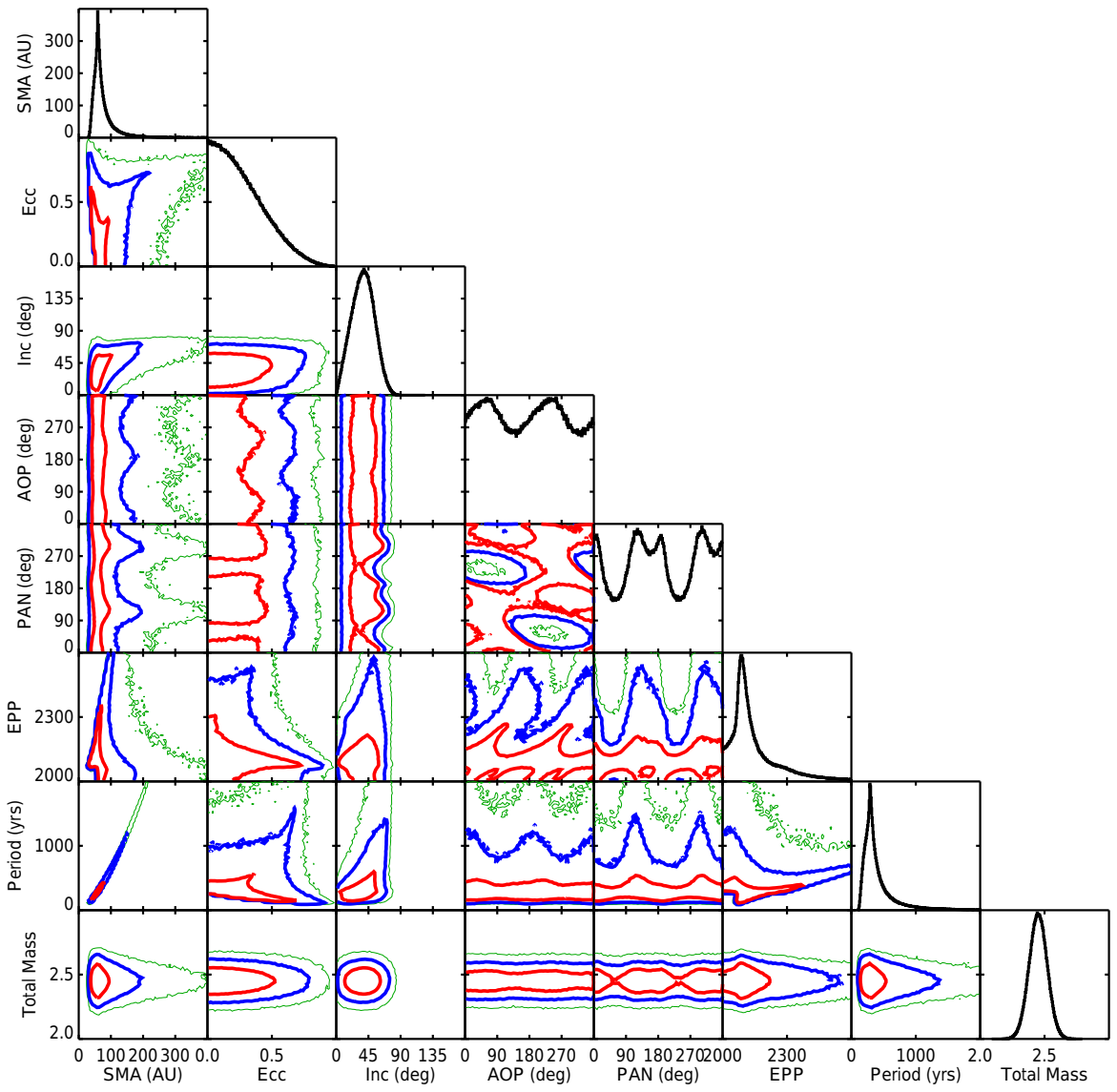

**Figure A2.** Posteriors from the OFTI orbit fit of HD 143811 AB b, based on three epochs of relative astrometry. Orbits with eccentricity ≲0.5 are generally preferred, with semimajor axis of ∼60 au and period of ∼300 yr. Red, blue, and green contours correspond to $1\sigma$, $2\sigma$, and $3\sigma$ confidence intervals, respectively.

## Appendix B
## GPI Spectrum

We include a table of the spectrum extracted from the 2016 April 30 GPI observation of HD 143811 AB b (Table B1).

**Table B1**
2016 GPI Spectrum of HD 143811 AB b

| Wavelength ($\mu$m) | Flux Ratio | Flux Ratio Error | Flux ($Wm^{-2}\mu m^{-1}$) | Flux Error ($Wm^{-2}\mu m^{-1}$) |
|---|---|---|---|---|
| 1.510 | 9.710e–06 | 4.99e–06 | 1.028e–17 | 6.18e–18 |
| 1.517 | 1.085e–05 | 5.66e–06 | 1.093e–17 | 6.94e–18 |
| 1.523 | 1.078e–05 | 6.40e–06 | 1.100e–17 | 7.75e–18 |
| 1.531 | 1.016e–05 | 5.27e–06 | 9.600e–18 | 6.29e–18 |
| 1.539 | 1.126e–05 | 4.81e–06 | 1.089e–17 | 5.65e–18 |
| 1.547 | 1.429e–05 | 4.26e–06 | 1.481e–17 | 4.94e–18 |
| 1.555 | 1.639e–05 | 3.00e–06 | 1.787e–17 | 3.42e–18 |
| 1.564 | 1.576e–05 | 2.86e–06 | 1.682e–17 | 3.20e–18 |
| 1.573 | 1.501e–05 | 3.80e–06 | 1.487e–17 | 4.18e–18 |
| 1.580 | 1.572e–05 | 4.44e–06 | 1.531e–17 | 4.81e–18 |
| 1.588 | 1.747e–05 | 3.80e–06 | 1.717e–17 | 4.06e–18 |
| 1.596 | 1.791e–05 | 3.56e–06 | 1.755e–17 | 3.76e–18 |
| 1.605 | 1.878e–05 | 4.24e–06 | 1.826e–17 | 4.41e–18 |
| 1.614 | 1.865e–05 | 4.34e–06 | 1.755e–17 | 4.43e–18 |
| 1.621 | 1.982e–05 | 3.89e–06 | 1.964e–17 | 3.91e–18 |
| 1.629 | 2.025e–05 | 3.60e–06 | 1.957e–17 | 3.56e–18 |
| 1.637 | 2.038e–05 | 3.77e–06 | 1.876e–17 | 3.66e–18 |
| 1.646 | 2.107e–05 | 3.88e–06 | 1.888e–17 | 3.69e–18 |
| 1.655 | 2.351e–05 | 4.05e–06 | 2.058e–17 | 3.79e–18 |
| 1.663 | 2.438e–05 | 4.64e–06 | 2.141e–17 | 4.27e–18 |
| 1.671 | 2.272e–05 | 4.77e–06 | 1.937e–17 | 4.31e–18 |
| 1.679 | 2.122e–05 | 4.05e–06 | 1.745e–17 | 3.59e–18 |
| 1.687 | 2.123e–05 | 4.14e–06 | 1.661e–17 | 3.62e–18 |
| 1.695 | 2.257e–05 | 4.22e–06 | 1.703e–17 | 3.64e–18 |
| 1.702 | 2.466e–05 | 4.49e–06 | 1.845e–17 | 3.82e–18 |
| 1.710 | 2.635e–05 | 5.57e–06 | 1.937e–17 | 4.68e–18 |
| 1.718 | 2.682e–05 | 6.47e–06 | 1.845e–17 | 5.32e–18 |
| 1.726 | 2.646e–05 | 6.44e–06 | 1.699e–17 | 5.17e–18 |
| 1.734 | 2.605e–05 | 6.32e–06 | 1.601e–17 | 4.97e–18 |
| 1.743 | 2.451e–05 | 5.33e–06 | 1.483e–17 | 4.13e–18 |
| 1.752 | 2.315e–05 | 6.09e–06 | 1.451e–17 | 4.66e–18 |
| 1.759 | 2.203e–05 | 6.84e–06 | 1.359e–17 | 5.19e–18 |
| 1.766 | 2.142e–05 | 5.46e–06 | 1.288e–17 | 4.10e–18 |
| 1.773 | 2.259e–05 | 5.06e–06 | 1.385e–17 | 3.76e–18 |

## ORCID iDs

Nathalie K. Jones https://orcid.org/0000-0002-4682-7797
Jason J. Wang https://orcid.org/0000-0003-0774-6502
Eric L. Nielsen https://orcid.org/0000-0001-6975-9056
Robert J. De Rosa https://orcid.org/0000-0002-4918-0247
Anne E. Peck https://orcid.org/0000-0003-2461-6881
William Roberson https://orcid.org/0009-0008-9687-1877
Jean-Baptiste Ruffio https://orcid.org/0000-0003-2233-4821
Jerry W. Xuan https://orcid.org/0000-0002-6618-1137
Bruce A. Macintosh https://orcid.org/0000-0003-1212-7538
S. Mark Ammons https://orcid.org/0000-0001-5172-7902
Vanessa P. Bailey https://orcid.org/0000-0002-5407-2806
Travis S. Barman https://orcid.org/0000-0002-7129-3002
Joanna Bulger https://orcid.org/0000-0003-4641-2003
Eugene Chiang https://orcid.org/0000-0002-6246-2310
Jeffrey K. Chilcote https://orcid.org/0000-0001-6305-7272
Gaspard Duchêne https://orcid.org/0000-0002-5092-6464
Thomas M. Esposito https://orcid.org/0000-0002-0792-3719
Michael P. Fitzgerald https://orcid.org/0000-0002-0176-8973
Katherine B. Follette https://orcid.org/0000-0002-7821-0695
Stephen Goodsell https://orcid.org/0000-0002-4144-5116
Alexandra Z. Greenbaum https://orcid.org/0000-0002-7162-8036
Pascale Hibon https://orcid.org/0000-0003-3726-5494
Patrick Ingraham https://orcid.org/0000-0003-3715-8138
Paul Kalas https://orcid.org/0000-0002-6221-5360
Quinn M. Konopacky https://orcid.org/0000-0002-9936-6285
Michael C. Liu https://orcid.org/0000-0003-2232-7664
Franck Marchis https://orcid.org/0000-0001-7016-7277
Christian Marois https://orcid.org/0000-0002-4164-4182
Brenda Matthews https://orcid.org/0000-0003-3017-9577
Dimitri Mawet https://orcid.org/0000-0002-8895-4735
Stanimir Metchev https://orcid.org/0000-0003-3050-8203
Maxwell A. Millar-Blanchaer https://orcid.org/0000-0001-6205-9233
Rebecca Oppenheimer https://orcid.org/0000-0001-7130-7681
Jenny Patience https://orcid.org/0000-0001-9004-803X
Marshall D. Perrin https://orcid.org/0000-0002-3191-8151
Lisa Poyneer https://orcid.org/0000-0002-4300-3426

Abhijith Rajan https://orcid.org/0000-0002-9246-5467
Julien Rameau https://orcid.org/0000-0003-0029-0258
Fredrik T. Rantakyrö https://orcid.org/0000-0002-9667-2244
Bin Ren https://orcid.org/0000-0003-1698-9696
Aniket Sanghi https://orcid.org/0000-0002-1838-4757
Dmitry Savransky https://orcid.org/0000-0002-8711-7206
Adam C. Schneider https://orcid.org/0000-0002-6294-5937
Anand Sivaramakrishnan https://orcid.org/0000-0003-1251-4124
Adam J. R. W. Smith https://orcid.org/0000-0002-9156-9651
Inseok Song https://orcid.org/0000-0002-5815-7372
Remi Soummer https://orcid.org/0000-0003-2753-2819
Sandrine Thomas https://orcid.org/0000-0002-9121-3436
Kimberly Ward-Duong https://orcid.org/0000-0002-4479-8291
Schuyler G. Wolff https://orcid.org/0000-0002-9977-8255

## References

Astropy Collaboration, Price-Whelan, A. M., Lim, P. L., et al. 2022, ApJ, 935, 167
Astropy Collaboration, Price-Whelan, A. M., Sipőcz, B. M., et al. 2018, AJ, 156, 123
Astropy Collaboration, Robitaille, T. P., Tollerud, E. J., et al. 2013, A&A, 558, A33
Bailey, V., Meshkat, T., Reiter, M., et al. 2014, ApJL, 780, L4
Baraffe, I., Chabrier, G., Barman, T. S., Allard, F., & Hauschildt, P. H. 2003, A&A, 402, 701
Bardalez Gagliuffi, D. C., Balmer, W. O., Pueyo, L., et al. 2025, ApJL, 988, L18
Best, W. M. J., Dupuy, T. J., Liu, M. C., et al. 2024, The UltracoolSheet: Photometry, Astrometry, Spectroscopy, and Multiplicity for 4000+ Ultracool Dwarfs and Imaged Exoplanets, v2.0.0, Zenodo, doi:10.5281/zenodo.10573247
Best, W. M. J., Liu, M. C., Magnier, E. A., & Dupuy, T. J. 2021, AJ, 161, 42
Best, W. M. J., Magnier, E. A., Liu, M. C., et al. 2018, ApJS, 234, 1
Blunt, S., Nielsen, E. L., De Rosa, R. J., et al. 2017, AJ, 153, 229
Bond, C. Z., Wizinowich, P., Chun, M., et al. 2018, SPIE, 10703, 107031Z
Brandt, T. D. 2021, ApJS, 254, 42
Bryan, M. L., Chiang, E., Morley, C. V., Mace, G. N., & Bowler, B. P. 2021, AJ, 162, 217
Castelli, F., & Kurucz, R. L. 2003, IAUS, 210, A20
Charnay, B., Bézard, B., Baudino, J. L., et al. 2018, ApJ, 854, 172
Chauvin, G., Desidera, S., Lagrange, A. M., et al. 2017, A&A, 605, L9
Chauvin, G., Gratton, R., Bonnefoy, M., et al. 2018, A&A, 617, A76
Chauvin, G., Lagrange, A. M., Dumas, C., et al. 2004, A&A, 425, L29
Chilcote, J., Konopacky, Q., Hamper, R., et al. 2024, SPIE, 13096, 1309699
Chilcote, J. K., Larkin, J. E., Maire, J., et al. 2012, SPIE, 8446, 84468W
Choi, J., Dotter, A., Conroy, C., et al. 2016, ApJ, 823, 102
Christiaens, V., Gonzalez, C., Farkas, R., et al. 2023, JOSS, 8, 4774
Currie, T., Daemgen, S., Debes, J., et al. 2014, ApJL, 780, L30
De Rosa, R. J., Esposito, T. M., Gibbs, A., et al. 2020a, SPIE, 11447, 114475A
De Rosa, R. J., Esposito, T. M., Hirsch, L. A., et al. 2019, AJ, 158, 225
De Rosa, R. J., Nielsen, E. L., Blunt, S. C., et al. 2015, ApJL, 814, L3
De Rosa, R. J., Nielsen, E. L., Wahhaj, Z., et al. 2023, A&A, 672, A94
De Rosa, R. J., Nguyen, M. M., Chilcote, J., et al. 2020b, JATIS, 6, 015006
de Zeeuw, P. T., Hoogerwerf, R., de Bruijne, J. H. J., Brown, A. G. A., & Blaauw, A. 1999, AJ, 117, 354
Deacon, N. R., Liu, M. C., Magnier, E. A., et al. 2014, ApJ, 792, 119
Delorme, P., Gagné, J., Girard, J. H., et al. 2013, A&A, 553, L5
Dotter, A. 2016, ApJS, 222, 8
Dupuy, T. J., & Kraus, A. L. 2013, Sci, 341, 1492
Dupuy, T. J., & Liu, M. C. 2012, ApJS, 201, 19
Esposito, T. M., Kalas, P., Fitzgerald, M. P., et al. 2020, AJ, 160, 24
Feroz, F., Hobson, M. P., & Bridges, M. 2009, MNRAS, 398, 1601
Foreman-Mackey, D., Hogg, D. W., Lang, D., & Goodman, J. 2013, PASP, 125, 306
Franson, K., Bowler, B. P., Zhou, Y., et al. 2023, ApJL, 950, L19
Gagné, J., Mamajek, E. E., Malo, L., et al. 2018, ApJ, 856, 23
Gaia Collaboration, Vallenari, A., Brown, A. G. A., et al. 2023, A&A, 674, A1
Galicher, R., Marois, C., Macintosh, B., Barman, T., & Konopacky, Q. 2011, ApJL, 739, L41
Galli, P. A. B., Joncour, I., & Moraux, E. 2018, MNRAS, 477, L50
Goldman, B., Marsat, S., Henning, T., Clemens, C., & Greiner, J. 2010, MNRAS, 405, 1140
Gomez Gonzalez, C. A., Wertz, O., Absil, O., et al. 2017, AJ, 154, 7
Grandjean, A., Lagrange, A. M., Meunier, N., et al. 2023, A&A, 669, A12
Greenbaum, A. Z., Pueyo, L., Ruffio, J.-B., et al. 2018, AJ, 155, 226
Harris, C. R., Millman, K. J., van der Walt, S. J., et al. 2020, Natur, 585, 357
Hinkley, S., Lacour, S., Marleau, G. D., et al. 2023, A&A, 671, L5
Hunter, J. D. 2007, CSE, 9, 90
Husser, T. O., Wende-von Berg, S., Dreizler, S., et al. 2013, A&A, 553, A6
Janson, M., Gratton, R., Rodet, L., et al. 2021, Natur, 600, 231
Kaufer, A., Stahl, O., Tubbesing, S., et al. 1999, Msngr, 95, 8
Konopacky, Q. M., Rameau, J., Duchêne, G., et al. 2016, ApJL, 829, L4
Konopacky, Q. M., Thomas, S. J., Macintosh, B. A., et al. 2014, SPIE, 9147, 914784
Koposov, S., Speagle, J., Barbary, K., et al. 2024, joshspeagle/dynesty: v2.1.4, Zenodo, doi:10.5281/zenodo.12537467
Kraus, A. L., Ireland, M. J., Cieza, L. A., et al. 2014, ApJ, 781, 20
Kraus, A. L., Ireland, M. J., Huber, D., Mann, A. W., & Dupuy, T. J. 2016, AJ, 152, 8
Kuzuhara, M., Tamura, M., Ishii, M., et al. 2011, AJ, 141, 119
Lagrange, A. M., Bonnefoy, M., Chauvin, G., et al. 2010, Sci, 329, 57
Lagrange, A. M., Gratadour, D., Chauvin, G., et al. 2009, A&A, 493, L21
Liu, M. C. 2004, Sci, 305, 1442
Liu, M. C., Dupuy, T. J., & Allers, K. N. 2016, ApJ, 833, 96
Luhman, K. L., & Esplin, T. L. 2020, AJ, 160, 44
Luhman, K. L., Tremblin, P., Birkmann, S. M., et al. 2023, ApJL, 949, L36
Macintosh, B., Graham, J. R., Barman, T., et al. 2015, Sci, 350, 64
Macintosh, B., Graham, J. R., Ingraham, P., et al. 2014, PNAS, 111, 12661
Marois, C., Doyon, R., Racine, R., & Nadeau, D. 2000, PASP, 112, 91
Marois, C., Lafrenière, D., Doyon, R., Macintosh, B., & Nadeau, D. 2006a, ApJ, 641, 556
Marois, C., Lafrenière, D., Macintosh, B., & Doyon, R. 2006b, ApJ, 647, 612
Marois, C., Lardière, O., Fitzsimmons, J., et al. 2024, SPIE, 13097, 1309708
Marois, C., Macintosh, B., Barman, T., et al. 2008, Sci, 322, 1348
Marois, C., Macintosh, B., & Véran, J.-P. 2010, SPIE, 7736, 77361J
Matthews, E. C., Carter, A. L., Pathak, P., et al. 2024, Natur, 633, 789
Mesa, D., Gratton, R., Kervella, P., et al. 2023, A&A, 672, A93
Millar-Blanchaer, M. A., Graham, J. R., Pueyo, L., et al. 2015, ApJ, 811, 18
Milli, J., Hibon, P., Christiaens, V., et al. 2017, A&A, 597, L2
Nasedkin, E., Mollière, P., Lacour, S., et al. 2024, A&A, 687, A298
Nguyen, M. M., De Rosa, R. J., & Kalas, P. 2021, AJ, 161, 22
Nielsen, E. L., De Rosa, R. J., Macintosh, B., et al. 2019, AJ, 158, 13
Nielsen, E. L., Liu, M. C., Wahhaj, Z., et al. 2013, ApJ, 776, 4
Nielsen, E. L., Rosa, R. J. D., Rameau, J., et al. 2017, AJ, 154, 218
Nowak, M., Lacour, S., Lagrange, A. M., et al. 2020, A&A, 642, L2
Ochsenbein, F., Bauer, P., & Marcout, J. 2000, A&AS, 143, 23
Pecaut, M. J., & Mamajek, E. E. 2016, MNRAS, 461, 794
Pecaut, M. J., Mamajek, E. E., & Bubar, E. J. 2012, ApJ, 746, 154
Peck, A. E., Roberson, W., Roberson, E. L., et al. 2025, ApJL, 995, L40
Perrin, M. D., Ingraham, P., Follette, K. B., et al. 2016, SPIE, 9908, 990837
Perrin, M. D., Maire, J., Ingraham, P., et al. 2014, SPIE, 9147, 91473J
Poyneer, L. A., Palmer, D. W., Macintosh, B., et al. 2016, ApOpt, 55, 323
Pueyo, L. 2016, ApJ, 824, 117
Rajan, A., Rameau, J., De Rosa, R. J., et al. 2017, AJ, 154, 10
Rameau, J., Chauvin, G., Lagrange, A.-M., et al. 2013, ApJL, 779, L26
Rodet, L., Beust, H., Bonnefoy, M., et al. 2017, A&A, 602, A12
Ruane, G., Ngo, H., Mawet, D., et al. 2019, AJ, 157, 118
Ruffio, J.-B., Macintosh, B., Wang, J. J., et al. 2017, ApJ, 842, 14
Sanghi, A., Liu, M. C., Best, W. M. J., et al. 2023, ApJ, 959, 63
Schneider, A. C., Munn, J. A., Vrba, F. J., et al. 2023, AJ, 166, 103
Service, M., Lu, J. R., Campbell, R., et al. 2016, PASP, 128, 095004
Sivaramakrishnan, A., & Oppenheimer, B. R. 2006, ApJ, 647, 620
Skilling, J. 2004, AIPC, 735, 395
Skilling, J. 2006, BayAn, 1, 833
Skrutskie, M. F., Cutri, R. M., Stiening, R., et al. 2006, AJ, 131, 1163
Snellen, I. A. G., Brandl, B. R., de Kok, R. J., et al. 2014, Natur, 509, 63
Soummer, R., Sivaramakrishnan, A., Pueyo, L., Macintosh, B., & Oppenheimer, B. R. 2011, ApJ, 729, 144
Speagle, J. S. 2020, MNRAS, 493, 3132
Squicciarini, V., Mazoyer, J., Wilkinson, C., et al. 2025, A&A, 702, L10
Stolker, T., Quanz, S. P., Todorov, K. O., et al. 2020, A&A, 635, A182
Stolker, T., Samland, M., Waters, L. B. F. M., et al. 2025, A&A, 700, A21

Thalmann, C., Desidera, S., Bonavita, M., et al. 2014, A&A, 572, A91
van Capelleveen, R. F., Kenworthy, M. A., Ginski, C., et al. 2025, arXiv:2508.18456
Walker, S., Liu, M., Mawet, D., et al. 2025, AJ, submitted
Wang, J. J., Ginzburg, S., Ren, B., et al. 2020, AJ, 159, 263
Wang, J. J., Graham, J. R., Dawson, R., et al. 2018a, AJ, 156, 192
Wang, J. J., Graham, J. R., Pueyo, L., et al. 2016, AJ, 152, 97
Wang, J. J., Perrin, M. D., Savransky, D., et al. 2018b, JATIS, 4, 018002
Wang, J. J., Rajan, A., Graham, J. R., et al. 2014, SPIE, 9147, 914755
Wang, J. J., Ruffio, J.-B., De Rosa, R. J., et al., 2015 Astrophysics Source Code Library, ascl:1506.001
Ward-Duong, K., Patience, J., Follette, K., et al. 2021, AJ, 161, 5
Wu, Y.-L., Bowler, B. P., Sheehan, P. D., et al. 2022, ApJL, 930, L3
Xuan, J. W., Hsu, C.-C., Finnerty, L., et al. 2024, ApJ, 970, 71
Xuan, W. J., Mawet, D., Ngo, H., et al. 2018, AJ, 156, 156
Zakhozhay, O. V., Launhardt, R., Müller, A., et al. 2022, A&A, 667, A63
Zhang, Z., Liu, M. C., Marley, M. S., Line, M. R., & Best, W. M. J. 2021, ApJ, 916, 53